\documentclass{article}
\def\beq{\begin{equation}}
\def\eeq{\end{equation}}
\def\bea{\begin{eqnarray}}
\def\eea{\end{eqnarray}}
\def\bq{\begin{quote}}
\def\eq{\end{quote}}

\def\ga{\left(}
\def\dr{\right)}

\def\rar{\rightarrow}

\def\la{\langle}
\def\ra{\rangle}
\def\nin{\noindent}
\def\ba{\begin{array}}
\def\ea{\end{array}}

\def\b{\bullet}

\begin{document}
\topmargin -1.5cm
\oddsidemargin +0.2cm
\evensidemargin -1.0cm
\pagestyle{empty}
\begin{flushright}
PM/97-50\\ 
\end{flushright}
\vspace*{5mm}
\begin{center}
\section*{Gluonia in QCD with massless quarks}
\vspace*{1.5cm}
{\bf Stephan Narison}
\\
\vspace{0.3cm}
Laboratoire de Physique Math\'ematique et Th\'{e}orique\\
UM2, Place Eug\`ene Bataillon\\
34095 - Montpellier Cedex 05, France\\
E-mail:
narison@lpm.univ-montp2.fr\\
\vspace*{2.5cm}
{\bf Abstract} \\ \end{center}
\vspace*{2mm}
\noindent
I briefly review \footnote{For more details and complete references, see: 
S.N., hep-ph/9612457 (1996) [Nucl. Phys. B (in press)] and QCD 97 International Euroconference,
Montpellier, hep-ph/9710281 (1997) [Nucl.Phys. B (Proc. Suppl.) (in press)].}
 the estimate of the gluonia
masses, decay and mixings in QCD with massless quarks from
QCD spectral sum rules and some low-energy theorems. The data in the
$0^{++}$ channel can be explained with some
maximal gluonium-quarkonium mixing schemes, which then suggest a large violation of the OZI rule,
similar to the case of the $\eta'$ in the $U(1)_A$ sector. 
\noindent
\vspace*{7cm}
%\rule[.1in]{15.0cm}{.002in}   
\begin{flushleft}
PM/97-50 \\
November 1997
\end{flushleft}
\vfill\eject

\pagestyle{plain}
\setcounter{page}{1}
\section{Introduction}
\nin
The possible existence of the gluon bound states 
(gluonia or glueballs) or/and of a gluon continuum is
one of the main consequences of the non-perturbative aspects
of QCD.
In this talk, I summarize our recent results on these topics.
\section {Gluonia masses from QCD spectral sum rules}
We shall consider the lowest-dimension gauge-invariant
currents $J_G$ built from two gluon fields with the 
quantum numbers of the $J^{PC}= 0^{++},~2^{++}$ and $0^{-+}$ 
gluonia. The former two enter 
the QCD energy-momentum tensor
$\theta_{\mu\nu}$, while the third one is the U(1)$_A$ axial anomaly. 
We shall also consider the
three-gluon current asssociated to the $0^{++}$ gluonia. 
 We shall work with the generic 
two-point correlator: 
$
\psi_G(q^2) \equiv i \int d^4x ~e^{iqx} \
\la 0\vert {\cal T}
J_G(x)
\ga J_G(0)\dr ^\dagger \vert 0 \ra ,
$
where its QCD expression
can be parametrized by the usual perturbative terms plus
the non-perturbative ones due to the vacuum
condensates in the Wilson expansion
\footnote{Higher dimension condensates, including the
ones due instanton--anti-instanton $(D=11)$ operators, can be neglected
at the sum rule optimization scale. UV renormalon and some eventual other
effects induced by the resummation of the QCD series, and not included in the OPE, 
are estimated from the last known term of the truncated series, as is done 
in the extraction of
$\alpha_s$ from $\tau$ decays.}. 
In the massless quark limit $m_i=0$, the dominant non-perturbative contribution 
is due to
the dimension-four gluonic condensate 
$\la\alpha_s G^2 \ra
\simeq (0.07\pm 0.01)~\mbox{GeV}^4$, recently estimated from the $e^+e^-$ data and
heavy-quark mass splittings 
\footnote{Throughout this paper, we shall use, for three active flavours,
the value of the QCD scale $
\Lambda= (375\pm 125)~\mbox{MeV}.$}.
We shall study the SVZ
Laplace sum rules:
\beq
{\cal L}_G(\tau)
= \int_{t_\leq}^{\infty} {dt}~\mbox{exp}(-t\tau)
~\frac{1}{\pi}~\mbox{Im} \psi_G(t),~~~{\mbox {and}}~~~
{\cal R}_G \equiv -\frac{d}{d\tau} \log {{\cal L}_G},
\eeq
where $t_\leq$ is the hadronic threshold. The latter,
 or its slight modification, is very useful, as it is equal to the 
resonance mass squared, in  
 the simple duality ansatz parametrization of the spectral function:
$
\frac{1}{\pi}\mbox{ Im}\psi_G(t)\simeq 2f^2_GM_G^4\delta(t-M^2_G)\delta(t-M^2_R)
 \ + \ 
 ``\mbox{QCD continuum}" \Theta (t-t_c),
$
where the
decay constant $f_G$ is analogous to $f_\pi=93.3$ MeV; 
$t_c$ is
the continuum threshold which is, like the 
sum rule variable $\tau$, an  a priori arbitrary 
parameter.  Our results in Table 1 satisfy
the $t_c$ and $\tau$ stability criteria, whilst the upper 
bounds
have been obtained from the minimum of ${\cal R}_G$ combined with its positivity.
Our results
show the mass hierarchy $M_S<M_P\approx M_T$, which suggests that the
scalar is the lightest gluonium state. However, 
the consistency of the different subtracted and unsubtracted
sum rules in the scalar sector requires
the existence of an additional lower mass and broad
$\sigma_B$-meson coupled strongly both to gluons and
to pairs of Goldstone bosons (a case
similar to the $\eta'$)\footnote{An analogous large 
violation of the OZI rule is also necessary for explaining the proton spin crisis.
These features are consequences of the $U(1)$ symmetry, which is not the case of the vector meson
$\phi$ of the $SU(3)_F$ symmetry.}. 
The effect of the $\sigma_B$-meson can be
missed in a one-resonance parametrization of the spectral function, and
in the present lattice quenched approximation. The values of $\sqrt{t_c}$,
which are about equal to the mass of the next radial excitations, indicate that the
mass splitting between the ground state and the radial excitations is much
smaller ($30\%$) than for the $\rho$ meson (about $70\%$), so
that one can expect rich gluonia spectra in the vicinity of 2.0--2.2 GeV. We also conclude
that:\\
$\b$ The $\zeta(2.2)$ is a good $2^{++}$ gluonium candidate because of its mass
(see Table 1) \footnote{The small quarkonium-gluonium (mass) mixing angle
allows us to expect that the observed meson mass is about the same as the one
in Table 1.} and small width in $\pi\pi$ ($\leq$ 100 MeV). However,
the associated value of $t_c$ can suggest that
the radial excitation state is also in the 2 GeV region, which
should stimulate further experimental searches.\\
$\b$ The $E/\iota$ (1.44) or other particles in this region is
too low for being the lowest pseudoscalar gluonium. One of these states is
likely to be the first radial excitation of the $\eta'$ because its coupling to the
gluonic current is weaker than the one of the $\eta'$ and of the
gluonium (see Table 1).
\section{Widths of the scalar gluonia}
$\b$ The hadronic widths can be estimated from the vertex:
$
V(q^2)=\la H_1|\theta^\mu_\mu|H_2\ra$, where: $q=p_1-p_2,~H\equiv \pi,\eta_1,\sigma_B
$, and
$B$ refers to the unmixed bare state. It satisfies the constraints $V(0)=M^2_H$
and, in the chiral limit, $V'(0)=1$. Saturating it with the three resonances $H\equiv\sigma_B,\sigma'_B$ and $G$, 
one obtains the first and second NV sum rules (Goldberger--Treiman-like relation):
\beq
\frac{1}{4}\sum_{S\equiv\sigma_B,\sigma'_B,G}g_{SHH}\sqrt{2}f_S \simeq 2M^2_H,~~~~~~~
\frac{1}{4}\sum_{S\equiv\sigma_B,\sigma'_B,G}g_{S\pi\pi}\sqrt{2}f_S/M^2_S=1.
\eeq
We identify the $G$ with the $G(1.5\sim 1.6)$ at GAMS (an almost pure gluonium
candidate), and the
$\sigma_B$ and $\sigma'_B$ (its radial excitation) with the broad resonance below
1 GeV and the $f_0(1.37)$. In this way, we obtain the predictions in Table 2,
showing 
the presence of gluons inside the wave 
functions of the broad $\sigma_B$ and $\sigma'_B$, which 
decay copiously into $\pi\pi$, signalling a large violation of the OZI rule in this channel. 
For the $G$ meson, we deduce 
$g_{G\eta\eta}\simeq \sin\theta_Pg_{G\eta\eta'}$ ($\theta_P=-18^\circ$ being the pseudoscalar mixing
angle), implying a ratio of widths of about 0.22
in agreement with the GAMS data $r\simeq 0.34\pm 0.13$, 
but suggest that the Crystal Barrel particle
having $r\approx 1$ is a mixing between this gluonium and other states.
We also expect that the $4\pi$ decay of the $G(1.6)$ 
are mainly $S$-waves initiated 
from the decay of pairs of $\sigma_B$.\\
%\subsection{$\gamma\gamma$ widths and $J/\psi\rar\gamma S$ decays}
%\nin
$\b$ The $\gamma\gamma$ widths of the $\sigma_B,~\sigma'_B$ and $G$
 can be obtained (Table 2) by identifying the 
Euler-Heisenberg box two gluons--two photons effective Lagrangian,
with the scalar-$\gamma\gamma$ Lagrangian,
while the {$J/\psi\rar\gamma S$ radiative decays}
can be estimated, using dispersion relation
techniques, to this effective interaction.
%\subsection{Hadronic widths from some low-energy theorems}
\section{``Mixing-ology'' for the decay widths of scalar mesons}
$\b$ We consider that the observed $f_0$ and $\sigma$
result from the two-component mixing of the $\sigma_B$ and $S_2\equiv
\frac{1}{\sqrt{2}}(\bar uu +\bar dd)$ unmixed bare states.
Using the prediction on
$\Gamma({\sigma_B}\rar\gamma\gamma),$
and the experimental width $\Gamma(f_0\rar\gamma\gamma)\approx 0.3~\mbox{keV}$,
one obtains a maximal mixing angle:
$\theta_S\approx (40\sim 45)^\circ,$
which indicates that
the $f_0$ and $\sigma$ have a large amount of gluons in
their wave functions \footnote{This situation is quite similar to the case of 
the $\eta'$
in the pseudoscalar channel (mass given by its gluon component, but 
strong coupling to quarkonia).}.  
Then, one can deduce (in units of GeV): 
\bea\label{coupe}
 g_{f_0\pi^+\pi^-}&\simeq& (0.1\sim 2.6),~~ g_{f_0K^+K^-}\simeq 
-(1.3\sim 4.1),
%\nnb\\
% g_{\sigma\pi^+\pi^-}&\simeq& ~g_{\sigma K^+K^-}\simeq (4\sim 5),
\eea
which can give a simple explanation of the exceptional property of the $f_0$ 
(strong coupling to $\bar KK$: $\pi\pi$ and 
$\bar KK$ data, and $\phi\rar \gamma f_0$ decay). \\
$\b$ The
$f_0(1.37),f_0(1.50)$ can be well described by a $3\times 3$ mixing scheme involving the
$\sigma'_B(1.37),S_3(\bar ss)(1.47)$ and $G(1.5)$ gluonium, and contains a large amount of
glue and $\bar ss$ components. The different widths are given in Table 3. The $f_J(1.71)$, if
confirmed to be a $0^{++}$, can be essentially composed by the radial excitation
$S'_3(\bar ss)$ due to its main decay into $\bar KK$.
\section{Conclusions and acknowledgements}
I have shown that QCD spectral sum rules plus some QCD low-energy theorems can
provide a plausible explanation of the complex gluonia spectra. \\
I thank the cern theory division
for its hospitality.
\begin{table*}[hbt]
%\begin{center}
% space before first and after last column: 1.5pc
% space between columns: 3.0pc (twice the above)
\setlength{\tabcolsep}{1.05pc}
% -----------------------------------------------------
% adapted from TeX book, p. 241
%\newlength{\digitwidth} \settowidth{\digitwidth}{\rm 0}
%\catcode`?=\active \def?{\kern\digitwidth}
% -----------------------------------------------------
\caption{ Unmixed gluonia masses and couplings from QSSR}
%\begin{tabular*}{\textwidth}{@{}l@{\extracolsep{\fill}}rrrrr}
\begin{tabular}{c c c c c c}
\hline 
%\hline
% & & & & &\\
$J^{PC}$&Name&{Mass [GeV]}&&$f_G$ [MeV]&$\sqrt{t_c}$ [GeV]\\
%&&{[GeV]}&&[MeV]&[GeV]\\
%   & & & & &\\
&&Estimate&Upper bound&&\\
%&&&&&\\
\hline 
%\hline
% & & & & & \\
$0^{++}$&$G$&$1.5\pm 0.2$&$2.16\pm 0.22$&$390\pm 145$&$2.0\sim 2.1$\\
&$\sigma_B$&1.00 (input)&&1000&\\
&$\sigma'_B$&1.37 (input)&&600&\\
&$3G$&3.1&&62&\\
%&&&&&\\
$2^{++}$&$T$&$2.0\pm 0.1$&$2.7\pm 0.4$&$80\pm 14$&$2.2$\\
%&&&&&\\
$0^{-+}$&$P$&$2.05\pm 0.19$&$2.34\pm 0.42$&$8\sim 17$&$2.2$\\
&$E/\iota$&1.44 (input)&&$ 7 ~:~J/\psi\rar\gamma\iota$&\\
%&&&&&\\
\hline 
%\hline
\end{tabular}
%\end{center}
%\begin{center}
%\begin{center}
% space before first and after last column: 1.5pc
% space between columns: 3.0pc (twice the above)
\setlength{\tabcolsep}{.6pc}
% -----------------------------------------------------
% adapted from TeX book, p. 241
%\newlength{\digitwidth} \settowidth{\digitwidth}{\rm 0}
%\catcode`?=\active \def?{\kern\digitwidth}
% -----------------------------------------------------
\caption{ Unmixed scalar gluonia and quarkonia decays}
\begin{tabular}{ c c c c c c c c}
\hline 
%\hline
% & & & & & & &\\ 
Name&Mass&$\pi^+\pi^-$&$K^+K^-$&$\eta\eta$
&$\eta\eta'$&$(4\pi)_S$&$\gamma\gamma$\\ 
& [GeV]& [GeV]&[MeV]&
[MeV]&[MeV]&[MeV]&[keV]\\ 

%&&&&&&\\
\hline 
%&&&&&&&\\
{Glue}&&&&&&&\\
%&&&&&&&\\
$\sigma_B$&$0.75\sim 1.0$&$ 0.2\sim 0.5$&$SU(3)$&$SU(3)$&&&$0.2\sim 0.3$\\
&(input)&&&&&\\
$\sigma'_B$&1.37&$0.5\sim 1.3$&$SU(3)$&$SU(3)$&&$43\sim 316$&$0.7\sim 1.0$\\
&(input)&&&&&(exp)&\\
$G$&1.5&$\approx 0$&$\approx 0$&$1\sim 2$&$5\sim 10$&$60\sim 138$&$0.2\sim 1.8$\\
%&&&&$(\frac{g_{G\eta\eta}}{g_{G\eta\eta'}}\simeq 
%\sin{\theta_P})$&&&\\
%&&&&&&&\\
{Quark}
&&&&&&&\\
%&&&&&&&\\
$S_2$&1.&0.12&$SU(3)$&$SU(3)$&&&0.67\\
$S'_2$&$1.3\approx \pi'$&$0.30\pm 0.15$&$SU(3)$&$SU(3)$&&&$4\pm 2$\\
$S_3$&$1.47\pm 0.04$&&$73\pm 27$&$15\pm 6$&&&$0.4\pm 0.04$\\
$S'_3$&$\approx 1.7$&&$112\pm 50$&$SU(3)$&&&$1.1\pm 0.5$\\
%&&&&&&&\\
\hline
\end{tabular}
%\end{center}
%\end{table*}
%\begin{table*}[hbt]
%\begin{center}
%\begin{center}
% space before first and after last column: 1.5pc
% space between columns: 3.0pc (twice the above)
\setlength{\tabcolsep}{.9pc}
% -----------------------------------------------------
% adapted from TeX book, p. 241
%\newlength{\digitwidth} \settowidth{\digitwidth}{\rm 0}
%\catcode`?=\active \def?{\kern\digitwidth}
% -----------------------------------------------------
\caption{ Predicted decays of the observed scalar mesons}
\begin{tabular}{ c c c c c c c}
\hline 
%\hline
%  & & & & & &\\ 
Name&$\pi^+\pi^-$&$K^+K^-$&$\eta\eta$
&$\eta\eta'$&$(4\pi^0)_S$&$\gamma\gamma$\\
&[MeV]&[MeV]&
[MeV]&[MeV]&[MeV]&[keV]\\  
%&&&&&&\\
\hline 
%&&&&&&\\
$f_0(0.98)$&$ 0.2\sim 134$&Eq. (\ref{coupe})&$$&&&$\approx 0.3$\\
&&&&&&(exp)\\
$\sigma(0.75\sim 1)$&$300\sim 700$&$SU(3)$&$SU(3)$&&$$&$0.2\sim 0.5$\\
&&&&&&\\
$f_0(1.37)$&$22\sim 48$&$\approx 0$&$\leq 1$&$\leq 2.5$&$150$&$\leq 2.2$\\
&&(exp)&&&&\\
$f_0(1.5)$&25&$3\sim 12$&$1\sim 2$&$\leq 1$&$68\sim 105$&$\leq 1.6$\\
&(exp)&&&&(exp)&\\
$f_J(1.71)$&$\approx 0$&$112\pm 50$&$SU(3)$&&$\approx 0$&$1.1\pm 0.5$\\
%&&&&&&\\
\hline
\end{tabular}
%\end{center}
\end{table*} 
\end{document}